\documentstyle[12pt]{article}

\catcode`\@=11
\long\def\@makefntext#1{
\protect\noindent \hbox to 3.2pt {\hskip-.9pt
$^{{\ninerm\@thefnmark}}$\hfil}#1\hfill}		

\def\@makefnmark{\hbox to 0pt{$^{\@thefnmark}$\hss}}  
	
\def\ps@myheadings{\let\@mkboth\@gobbletwo
\def\@oddhead{\hbox{}
\rightmark\hfil\ninerm\thepage}
\def\@oddfoot{}\def\@evenhead{\ninerm\thepage\hfil
\leftmark\hbox{}}\def\@evenfoot{}
\def\sectionmark##1{}\def\subsectionmark##1{}}

\renewcommand{\thefootnote}{\fnsymbol{footnote}}

\newcounter{sectionc}\newcounter{subsectionc}\newcounter{subsubsectionc}
\renewcommand{\section}[1] {\vspace*{0.6cm}\addtocounter{sectionc}{1}
\setcounter{subsectionc}{0}\setcounter{subsubsectionc}{0}\noindent
	{\normalsize\bf\thesectionc. #1}\par\vspace*{0.4cm}}
\renewcommand{\subsection}[1] {\vspace*{0.6cm}\addtocounter{subsectionc}{1}
	\setcounter{subsubsectionc}{0}\noindent
	{\normalsize\it\thesectionc.\thesubsectionc. #1}\par\vspace*{0.4cm}}
\renewcommand{\subsubsection}[1]
{\vspace*{0.6cm}\addtocounter{subsubsectionc}{1}
	\noindent  
{\normalsize\rm\thesectionc.\thesubsectionc.\thesubsubsectionc.
	#1}\par\vspace*{0.4cm}}

\newcounter{appendixc}
\newcounter{subappendixc}[appendixc]
\newcounter{subsubappendixc}[subappendixc]

\renewcommand{\appendix}[1] {\vspace*{0.6cm}
        \refstepcounter{appendixc}
        \setcounter{figure}{0}
        \setcounter{table}{0}
        \setcounter{equation}{0}
        \renewcommand{\thefigure}{\Alph{appendixc}.\arabic{figure}}
        \renewcommand{\thetable}{\Alph{appendixc}.\arabic{table}}
        \renewcommand{\theappendixc}{\Alph{appendixc}}
        \renewcommand{\theequation}{\Alph{appendixc}.\arabic{equation}}
        \noindent{\bf Appendix \theappendixc #1}\par\vspace*{0.4cm}}

\def\abstracts#1{{
	\centering{\begin{minipage}{12.2truecm}\footnotesize\baselineskip=12pt\ 
noindent
	\centerline{\footnotesize ABSTRACT}\vspace*{0.3cm}
	\parindent=0pt #1
	\end{minipage}}\par}}

\renewenvironment{thebibliography}[1]
	{\begin{list}{\arabic{enumi}.}
	{\usecounter{enumi}\setlength{\parsep}{0pt}
\setlength{\leftmargin 1.25cm}{\rightmargin 0pt}
	 \setlength{\itemsep}{0pt} \settowidth
	{\labelwidth}{#1.}\sloppy}}{\end{list}}

\newcounter{itemlistc}
\newcounter{romanlistc}
\newcounter{alphlistc}
\newcounter{arabiclistc}

\newcommand{\fcaption}[1]{
        \refstepcounter{figure}
        \setbox\@tempboxa = \hbox{\footnotesize Fig.~\thefigure. #1}
        \ifdim \wd\@tempboxa > 6in
           {\begin{center}
        \parbox{6in}{\footnotesize\baselineskip=12pt Fig.~\thefigure. #1}
            \end{center}}
        \else
             {\begin{center}
             {\footnotesize Fig.~\thefigure. #1}
              \end{center}}
        \fi}

\newcommand{\tcaption}[1]{
        \refstepcounter{table}
        \setbox\@tempboxa = \hbox{\footnotesize Table~\thetable. #1}
        \ifdim \wd\@tempboxa > 6in
           {\begin{center}
        \parbox{6in}{\footnotesize\baselineskip=12pt Table~\thetable. #1}
            \end{center}}
        \else
             {\begin{center}
             {\footnotesize Table~\thetable. #1}
              \end{center}}
        \fi}

\def\@citex[#1]#2{\if@filesw\immediate\write\@auxout
	{\string\citation{#2}}\fi
\def\@citea{}\@cite{\@for\@citeb:=#2\do
	{\@citea\def\@citea{,}\@ifundefined
	{b@\@citeb}{{\bf ?}\@warning
	{Citation `\@citeb' on page \thepage \space undefined}}
	{\csname b@\@citeb\endcsname}}}{#1}}

\newif\if@cghi
\def\cite{\@cghitrue\@ifnextchar [{\@tempswatrue
	\@citex}{\@tempswafalse\@citex[]}}
\def\citelow{\@cghifalse\@ifnextchar [{\@tempswatrue
	\@citex}{\@tempswafalse\@citex[]}}
\def\@cite#1#2{{$\null^{#1}$\if@tempswa\typeout
	{IJCGA warning: optional citation argument
	ignored: `#2'} \fi}}

 1
 1
 1

\font\ninerm=cmr9

\def\av#1{\langle #1 \rangle}

\def\expect#1{{\rm E}\left\{ #1 \right\}}
\def\td{\tau_{\rm dep}}

\begin{document}
\rightline{\footnotesize OUCMT-96-8}
\vspace*{1cm}
\centerline{\normalsize\bf STATISTICAL DEPENDENCE ANALYSIS}
\baselineskip=22pt

\centerline{\footnotesize MACOTO KIKUCHI}
\baselineskip=13pt
\centerline{\footnotesize\it Department of Physics, Osaka University,
Toyonaka 560, Japan}
\baselineskip=12pt
\centerline{\footnotesize E-mail: kikuchi@phys.sci.osaka-u.ac.jp}
\vspace*{0.3cm}
\centerline{\footnotesize NOBUYASU ITO}
\baselineskip=13pt
\centerline{\footnotesize\it Department of Applied Physics,
The University of Tokyo, Tokyo 113, Japan}
\baselineskip=12pt
\centerline{\footnotesize E-mail: ito@catalyst.t.u-tokyo.ac.jp}
\vspace*{0.3cm}
\centerline{\footnotesize and}
\vspace*{0.3cm}
\centerline{\footnotesize YUTAKA OKABE}
\baselineskip=13pt
\centerline{\footnotesize\it Department of  Physics,
Tokyo Metropolitan University,
Hachioji 192-03, Japan}

\vspace*{0.9cm}
\abstracts{
We review our recent studies on the dynamical correlations in MC simulations  
from the view point of the statistical dependence.
Attentions are paid to the reduction of the statistical degrees of freedom for  
correlated data.
Possible biases on several cumulants, such as the susceptibility and the  
Binder number due to finite MC length are discussed.
A new method for calculating the equilibrium relaxation time from the analysis  
of the statistical dependence is presented.
We apply it to the critical dynamics of the Ising model to estimate the  
dynamical critical exponent accurately.}

\normalsize\baselineskip=15pt
\setcounter{footnote}{0}
\renewcommand{\thefootnote}{\alph{footnote}}

\section{Introduction}

Thanks to the very high performance of the computers today,
we are now able to calculate thermodynamic quantities highly precisely using  
Monte Carlo (MC) simulations.
At the same time,
importance of proper estimation of the statistical errors has widely been
recognized.
Since we can perform only MC runs of finite length,
statistical errors are inevitable.
Suppose we have a number of statistically independent data;
then analysis of the statistical errors are just a simple excercise of  
elementary statistics.
Difficulty arises, however, when we deal with data obtained by MC simulations;
since the MC simulations are dynamical methods, the data are
dynamically correlated and thus they are no longer statistically independent
with each other.

It is still not so difficult to estimate statistical errors when we deal with
such quantities as
the energy and the magnetization of classical spin systems and their higher  
moments.
In fact, their unbiased estimators are just the average of the measured values  
at each MC step.
The simplest unambiguous way for their error analysis is based on several  
independent MC runs of the same length.
When a number of independent averages are available,
then we can directly apply the elementary method of error estimation.
For any other quantity, however, there is no guarantee in general that this  
method works properly.
We can certainly get correct errors {\em around the estimates} by applying the  
error propagation law;
But the estimates themselves may be suffered from some biases due to the  
finite length of the MC run.
For example,
estimators for cumulants are usually biased, although they are consistent.
We can even estimate the magnitude of such bias, provided we know how many  
independent data we have.
Problem is that what we need is not just the number of actual measurments.
we rather need the number of {\it statistically independent} data.
In fact, data taken in a shorter interval than the relaxation time $\tau$
are strongly correlated with each other, so that the effective number of  
available statistically independent data
are less (sometimes much less) than the number of the actual measurements.
Intuitively speaking, the elementary interval between the
adjacent independent measurements would be $2\tau$ as was discusses by
M\"uller-Krumbhaar and Binder some twenty years ago\cite{MB73}.
Then, the number of independent data is estimated as $n/2\tau$, where
$n$ is the total number of measurements made in each run.
Based on this consideration,
effect of the biases on the susceptibility and the specific heat -- both are  
second-order cumulants --  due to finite MC length was discussed by Ferrenberg  
et al.\cite{FLB91}
Quite recently, we showed, in somewhat different context, that the independent  
data are more
accurately counted by introducing a new scale of time, the statistical  
dependence time, $\td$.\cite{KI93,KIO}
In contrast with the conventional reasoning where the number of independent  
data are simply proportional to the number of measurements,
we found that they relates with each other nontrivially.

Statistical errors and biases are not simply undesirable.
Rather, they offer useful informations on dynamical correlations,
because the dynamical correlation is just a different aspect of
the statistical dependence.
Then we can formulate yet another approach for studying dynamics using  
knowlege of the statistical errors and the biases;
we may call it {\em statistical dependence analysis} approach.
Recently, we proposed a new method for calculating $\tau$
from a ratio of equilibrium averages
of the susceptibility and the statistical error.\cite{KI93}
This method no longer requires calculations of time-displaced correlation
functions;
Thus its largest advantage over the conventional methods is that
it enables us to make unambiguous statistical analysis,
because we can simply follow the standard methods used for
static quantities.
We have applied this method to the MC simulations of the two- and  
three-dimensional Ising model, and
calculated the dynamical critical exponent $z$ highly accurately.

\section{The Statistical Dependence and dynamical correlation}

First, we discuss the relation among the susceptibility (second-order  
cumulants, in general), the relaxation time and the statistical error.
In the course of the discussion, we will examine the method for counting the  
number of the independent data properly.

Suppose we make $N$ identical MC simulations which differ only in the random  
number sequences.
Random number sequences should be independent from run to run, in order mutual  
statistical independence of these simulations to be guaranteed.
We make $n$ measurements of a quantity $Q$ in each run.
Let us define two types of averages as follows:
(1)The average over the data taken in a single run,
$ \av Q _\alpha \equiv  {1 \over n}\sum_{k=1}^{n} Q_\alpha(k)$, and
(2)The average over all the independent runs,
$ \overline{\av Q} \equiv {1 \over N}\sum_{\alpha =1}^{N} \av Q _\alpha$,
where $Q_\alpha(k)$ denotes the value of the quantity $Q$ at the $k$-th
measurement in the $\alpha$-th run, with $k=1$, $2$, $\cdots$, $n$ and
$\alpha =1$, $2$, $\cdots$, $N$.

Elementary statistics says that
when all the $n$ data in a run are statistically independent,
then the statistical degrees of freedom (DOF) also is $n$ and
the expectation value for the variance of $\av Q$
(we omitted the suffix $\alpha$ here by an apparent reason) is simply given as

\begin{equation}
\expect{(\av Q - \mu)^2} = \frac{\sigma^2}{n}, \label{eq:indep}
\end{equation}
where $\mu \equiv \expect Q$ and
$\sigma^2 \equiv \expect{Q^2} - \mu^2$ are the expectation values of $Q$
and of the variance of $Q$, respectively.
In the present situation, however,
the data in a run are dynamically correlated with each other
rather than they are independent.
The reduction in the DOF due to these correlations
can be discussed in terms of the time-displaced
correlation function of $Q$.
The expectation value for the variance of $\av Q$ can be written as

\begin{equation}
\expect{(\av Q - \mu)^2} =
 {1 \over {n^2}} \sum_{k, l = 1}^{n} \left( \expect{m(k) m(l)} - \mu^2\right)
 = {{\sigma^2} \over {n^2}} \sum_{k, l = 1}^{n} \expect{C(k,l)}.
\label{eq:expect1}
\end{equation}
where $C(k,l)$ is the (normalized) time-displaced correlation function of
the fluctuation in $Q$ for ``time'' $k$ and $l$.
By comparing Eq.~(\ref{eq:expect1}) with Eq.~(\ref{eq:indep}),
we find that the DOF is reduced from $n$ to
$n_{red} \equiv n^2 / \sum_{k, l = 1}^{n} \expect{C(k,l)}$.
Thus the reduction factor $n_{red}/n$ depends highly nontrivially on both $n$  
and $\tau$.

Let us introduce a new quantity, $\td$ with the dimension of time as

\begin{equation}
\td \equiv \frac{1}{2n} \sum_{k, l = 1}^{n} \expect{C(k,l)}.
\end{equation}
Then we can put Eq.~(\ref{eq:expect1}) in a similar form as
Eq.~(\ref{eq:indep}):

\begin{equation}
 \expect{(\av Q - \mu)^2} = {{2 \td}\over n} \sigma^2, \label{eq:reduc}
\end{equation}
The DOF is thus reduced by a factor $1/2\td$ due to the dynamical  
correlations.
In other words,
we can interprete $2\td$ as the mean interval of
successive statistically independent measurements.
It may be somewhat surprising result, because this interval depends not only  
on
the relaxation time $\tau$ but also on the total number of the measurements
$n$;
Namely, this interval is determined only after all the measurements have been  
made, even if we know $\tau$ beforehand.

Equation~(\ref{eq:reduc}) tells us how we can estimate $\td$ and, as a result,  
the reduced DOF by simulations:
The unbiased estimator $(\delta Q)^2$ for $\expect{(\av Q - \mu)^2}$
is calculated from $N$ independent runs as
$(\delta Q)^2
= {N\over{N-1}}\left( \overline{\av Q^2} - \overline{\av Q}^2 \right)
$
with $|\delta Q|/\sqrt{N}$ being the
statistical error in $Q$.
The variance $\sigma^2$, on the other hand, is calculated as
the susceptibility $\chi_Q$ associated to $Q$,
$
\chi_Q = \overline{\av{Q^2}} - \overline{\av Q}^2$.
Thus we get the estimator of $\td$ as follows:

\begin{equation}
\td = {{n (\delta Q)^2}\over{2 \chi_Q}}
   = {{nN}\over {2(N-1)}}
     {{\overline{\av Q^2} - \overline{\av Q}^2 }\over
      {\overline{\av{Q^2}} - \overline{\av Q}^2 }}, \label{eq:tdest}
\end{equation}
where the r.h.s. can be calculated by simulations.
Let us see how $\td$ behaves.
For simplicity, we assume a single exponential relaxation for
the correlation function, that is, we put
$\expect{C(k,l)} = \exp(-|k-l|/\tau)$.
If we approximate the summation in Eq.~(\ref{eq:expect1}) by an integration in  
the range $[0,\infty)$,
we reproduce the result obtained
by M\"uller-Krumbhaar and Binder\cite{MB73},
which implies that the reduced DOF is given as
$n_{red} = n/2\tau$.
From the derivation, however, this expression is valid
only if two conditions $n \ll \tau$ and $\tau \ll 1$ are satisfied,
that is, in the long-time and the long-relaxation-time limit.

We can go a little further.
The summation in Eq.~(\ref{eq:expect1}) can be calculated explicitly
without assuming the long-time nor the long-relaxation-time limit.
And we get
\begin{equation}
\td = {1\over 2}\left[{{1+\Lambda}\over{1-\Lambda}}
      - {{2\Lambda (1-\Lambda ^n)}\over{n(1-\Lambda)^2}}\right], \label{eq:td}
\end{equation}
where $\Lambda \equiv e^{-1/\tau}$.
It can easily be confirmed that the reduced DOF, $n_{red}=n/2\td$,
varies smoothly from 1 for $n=1$ to $n/2\tau$ for $n\rightarrow\infty$
as it should do.
Moreover, we can use Eq.~(\ref{eq:td}) for estimating $\tau$, once $\td$ is  
estimated by the simulations.

\section{On Unbiased Estimators of Cumulants}

As we have mentioned in the introduction, estimators for cumulants of any  
order are, in principle, biased.
For example,
it is well known that the unbiased estimator for the variance,
that is, the second-order cumulant,
from $n$ {\em independent} measurements is given as

\begin{equation}
\chi_Q = {n\over{n-1}}\left( \av Q^2 - \av Q^2 \right).
\label{eq:cumulant}
\end{equation}
The factor $n/(n-1)$ appears here for correction of the bias
due to the finite number of the measurements.
Therefore, the estimators for susceptibilities
calculated from the fluctuations in general
should also be corrected by the above factor.
Otherwise, the calculated susceptibility will be systematically
underestimated.
Importance of correcting these systematic errors on susceptibilities was
discussed in ref.~2.
In actual simulations, the measurements are not independent
of each other, as has been discussed so far.
Therefore, we can not simply take $n/(n-1)$ as the correction factor.
In ref.~2, it is argued that the susceptibility estimated from
$n$ measurements $\chi_Q(n)$
and its true expectation value $\chi_Q(\infty)$ is related as

\begin{equation}
\chi_Q(n) = \chi_Q(\infty ) \left(1-\frac{2\tau+1}{n} \right).  
\label{eq:tscale}
\end{equation}
This form coincides with the actual MC data for rather large $n$.
It is not surprising that eq.~(\ref{eq:tscale}) deviates from the actual MC  
data for smaller $n$, because $n/(2\tau+1)$ was used as the DOF, which, as was  
discussed in the previous section, is valid only in the long-time limit.

As we have shown, the proper form of DOF is $n/2\td$ rather than $n/2\tau$.
Therefore, Eq.(\ref{eq:tscale}) should be modified as

\begin{equation}
\chi_Q(n) = \chi_Q(\infty ) \left(1-\frac{2\td}{n} \right). \label{eq:tscale2}
\end{equation}
For examining this equation, we made MC simulations of three-dimensional Ising  
model at the criticality taking several different lengths of the run.
The system size is $16^3$.
Figure~\ref{fig:sus} shows $n$ dependence of the magnetic susceptibility  
$\chi$ calculated as the second-order cumulant.
\begin{figure}[tbp]
\vspace*{13pt}
\leftline{\hfill\vbox{\hrule width 5cm height0.001pt}\hfill}
\vspace*{1.4truein}             
\leftline{\hfill\vbox{\hrule width 5cm height0.001pt}\hfill}
\fcaption{$n$ dependence of the magnetic susceptibility for
$16^3$ 3D Ising model. The solid line indicate Eq.~(\ref{eq:tscale2}).}
\label{fig:sus}
\end{figure}
The solid line is the expected behavior from Eq.~(\ref{eq:tscale2});
$\chi_Q(\infty )$ and $\tau$ were estimated from a longer run,
and then $\td$ was estimated using Eq.~(\ref{eq:td}).
We can clearly see the systematic deviation of the shourt-run values of $\chi$  
from the long-time average.
Moreover, this systematic behavior is really expressed by  
Eq.~(\ref{eq:tscale2})
even for rather short simulations.
By making the similar plot for the specific heat $C$,
we found that that $C$ also behaves as expected.
Thus the validity of eq.~(\ref{eq:tscale2}) has been verified.

So far, we have examined the bias on the second-order cumulants.
The origin of the bias is attributed to the fact that
the second-order cumulants are calculated as a combination of the estimators  
of the first-order and the second-order moments,
and thus are calculated only after the averages of all the involved moments  
are taken.
Another important quantity frequently used in study of phase transitions is  
also in this category:
that is, the Binder number.
The relating cumulant to the Binder number is a simplified version of the  
fourth-order cumulant, $3\av{Q^2}^2 - \av{Q^4}$.
It is easily confirmed that its unbiased estimator is given as follows:

\begin{equation}
3\expect{\av{Q^2}^2} -\expect{\av{Q^4}}
= 3\frac{n-1}{n}(\sigma^2 + \mu^2)^2 + \frac{3-n}{n}\mu_4, \label{eq:c4est}
\end{equation}
where $\mu_4 \equiv \expect{Q^4}$.
By dividing it by $\expect{\av{Q^2}^2}$ we get

\begin{equation}
\frac{\expect{\av{Q^4}}}{\expect{\av{Q^2}^2}}
= \frac{R_{\infty}}{1+\frac{1}{n}(R_{\infty}-1)}, \label{eq:ratioest}
\end{equation}
where $R_{\infty}$ is the true value for the ratio of the moments:
\begin{equation}
R_{\infty} \equiv \frac{\expect{Q^4}}{\expect{Q^2}^2}
= \frac{\mu_4}{(\sigma^2 + \mu^2)^2}.  \label{eq:rtrue}
\end{equation}
It, however, is not what we can use for deriving the unbiased estmator for the  
Binder parameter;
while it is the ratio of the expectation values of the moments,
what we really need is the the expectation value of the ratio,
$\expect{\av{Q^4}/\av{Q^2}^2}.$
Only if the correlation between the denominator and the numerator can be
ignored, we can use Eq.~(\ref{eq:ratioest}) as its approximant.
That is certainly not true in general, of course.
But, in any case, the bias of $O(1/n)$ is expected for the Binder number.

In Fig.~\ref{fig:sbin}, $n$ dependence of the ratio $\av{m^4}/\av{m^2}^2$ for  
the
magnetization moments is plotted.
\begin{figure}[tbp]
\vspace*{13pt}
\leftline{\hfill\vbox{\hrule width 5cm height0.001pt}\hfill}
\vspace*{1.4truein}             
\leftline{\hfill\vbox{\hrule width 5cm height0.001pt}\hfill}
\fcaption{$n$ dependence of $\av{m^4}/\av{m^2}^2$ for
$16^3$ 3D Ising model.}
\label{fig:sbin}
\end{figure}
The effect of the bias is clearly seen.
But Eq.~(\ref{eq:ratioest}) cannot reproduce this behavior.
The similar plot for the energy, $\av{E^4}/\av{E^2}^2$ is shown in  
Fig.~\ref{fig:ebin}.
The solid line indicate Eq.~(\ref{eq:ratioest}).
\begin{figure}[tbp]
\vspace*{13pt}
\leftline{\hfill\vbox{\hrule width 5cm height0.001pt}\hfill}
\vspace*{1.4truein}             
\leftline{\hfill\vbox{\hrule width 5cm height0.001pt}\hfill}
\fcaption{$n$ dependence of $\av{E^4}/\av{E^2}^2$ for
$16^3$ 3D Ising model. The solid line indicate Eq.~(\ref{eq:ratioest}).}
\label{fig:ebin}
\end{figure}
In contrast with the case of the magnetization,
we see that the behavior of the ratio is well approximated by  
Eq.~(\ref{eq:ratioest}).
This difference between two Bender numbers may be attributed to the fact that
the Binder number was originally derived from a simplified version of the  
fourth-order cumulant, with the first- and the third-order moments set to be  
zero beforehand.
For the magnetization, this simplification is meaningful, since the  
expectation values for all the odd-order moments vanish due to the  
time-reversal symmetry.
The energy, on the contrary, does not have this property.
Therefore, the energy Binder number has only a vague relation with the  
fourth-order cumulant.
That may be the reason why the behavior of the energy Binder number agrees  
with Eq.~(\ref{eq:ratioest}), while the magnetization Binder number does not.
In any case, their estimators approaches the true values slowly with $n$.

For the quantities we have dealt with, that is, the second-order cumulants and  
the Binder number,
we can easily realize that their estimator should be biased.
Suppose we have only one measurement, namely, take $n\rightarrow 1$;
their estimators give just trivial values in this extreme limit:
0 for the second-order cumulants and $2/3$ for the Binder number
irrespective of the measured values of the magnetization or the energy.
Therefore, it is obvious that the values of these estimators vary from these  
trivial values to the true values as  $n$ increases.

\section{Estimation of the Dynamical Critical Exponent}
As we have discussed, we can estimate the relaxation time $\tau$ using the  
averages of the statistical error and the susceptibility.
Since only imformations of {\it static} averages are used in this calculation,
it allows us of an unambiguous error estimation for $\tau$.
As a result, we can obtain a highly reliable estimate of $\tau$.
Thus it is a suitable method especially for calculating the dynamical critical  
exponent $z$.
We have applied this method to two- and three-dimensional Ising models at  
their criticality, and obtained very accurate estimates of $z$.
In what follows, we present the results briefly.
Details of the calculations are found in ref.~3.

Figure~\ref{fig:tau} shows the log-log plot of the relaxation time $\tau$  
against the linear dimension $L$ of the three-dimensional Ising model.
\begin{figure}[tbp]
\vspace*{13pt}
\leftline{\hfill\vbox{\hrule width 5cm height0.001pt}\hfill}
\vspace*{1.4truein}             
\leftline{\hfill\vbox{\hrule width 5cm height0.001pt}\hfill}
\fcaption{The finite-size scaling plot of $\tau$ for
3D Ising model. The slope of the line gives $z = 2.03 \pm 0.01$.}
\label{fig:tau}
\end{figure}
We estimated $\tau$ from magnetic susceptibility and the statistical error of  
the magnetization using Eq.~(\ref{eq:td}).
The data fit very well to a straight line.
According to the dynamical finite-size scaling theory,
the slope of the plot gives $z$.
After careful statistical analyses, we have got $z = 2.03 \pm 0.01$.
Seeing the statistical error,
the value of $z$ we obtained is the most accurate one
proposed so far, as far as we know.
By making the similar finite-size scaling analysis, we have got
$z=2.173\pm 0.016$ for two-dimentional Ising model.
The present statistical error is also
one of the smallest ones among the studies so far.
Most importantly,
we did not use any special technique for estimating the statistical error in  
$z$;
We just followed the standard error analysis methods used for
static quantities.

\bigskip
\bigskip
\noindent{\bf Acknowledgements:}
The work was partially supported by a Grant-in-Aid
for Scientific Research on Priority Areas, ``Computational Physics
as a New Frontier in Condensed Matter Research'', from the Ministry
of Education, Science and Culture, Japan.

\end{document}

\bibitem{CBS} C.~K.~Chakrabarti, H.~G.~Baumg\"artel, and D.~Stauffer:
Z.~Phys. {\bf B44} (1981) 333.
\bibitem{2)} M.~C.~Yalabik and J.~D.~Gunton: Phys.~Rev. {\bf B25} (1982)
534.
\bibitem{3)} R.~B.~Pearson, J.~L.~Richardson, and D.~Toussaint:
Phys.~Rev. {\bf 31} (1985) 4472.
\bibitem{12)} C.~Kalle: J.~Phys. {\bf A17} (1984) L801.
\bibitem{7)} M.~Kikuchi and Y.~Okabe: J.~Phys.~Soc.~Jpn.
{\bf 55} (1986) 1359.
\bibitem{8)} D.~Stauffer: Physica {\bf A186} (1992) 197.
\bibitem{9)} G.~A.~Kohring and D.~Stauffer: Int.~J.~Mod.~Phys.
{\bf C3} (1992) 1165.
\bibitem{I3D93} N.~Ito: Physica {\bf A192} (1993) 604.

\bibitem{TSC81} J.~Tobochnik, S.~Sarker, and R.~Cordery:
Phys.~Rev.~Lett. {\bf 46} (1981) 1417.
\bibitem{T82} H.~Takano: Prog.~Theor.~Phys. {\bf 68} (1982) 493.
\bibitem{KDL82} S.~L.~Katz, J.~D.~Gunton, and C.~P.~Liu:
Phys.~Rev. {\bf B25} (1982) 6008.
\bibitem{JMS83} N.~Jan, L.~L.~Moseley, and D.~Stauffer:
J.~Stat.~Phys. {\bf 33} (1983) 1.
\bibitem{W85} J.~K.~Williams: J.~Phys. {\bf A18} (1985) 49.
\bibitem{MT85} S.~Miyashita and H.~Takano:
Prog.~Theor.~Phys. {\bf 73} (1985) 1122.
\bibitem{TL87} S.~Tang and D.~P.~Landau: Phys.~Rev. {\bf B36} (1987) 567.
\bibitem{ITS87} N.~Ito, M.~Taiji, and M.~Suzuki: J.~Phys.~Soc.~Jpn.
{\bf 56} (1987) 4218.

\bibitem{I2D93} N.~Ito: Physica {\bf A196} (1993) 591.


\begin{thebibliography}{99}
\bibitem{MB73} H.~M\"uller-Krumbhaar and K.~Binder: J.~Stat.~Phys. {\bf 8}
(1973) 1.
\bibitem{FLB91} A.~M.~Ferrenberg, D.~P.~Landau, and K.~Binder:
J.~Stat.~Phys. {\bf 63} (1991) 867.
\bibitem{KI93} M.~Kikuchi and N.~Ito: J.~Phys.~Soc.~Jpn. {\bf 62} (1993) 3052.
\bibitem{KIO}  M.~Kikuchi, N.~Ito and Y.~Okabe:
in {\sl Computer Simulations in Condensed Matter Physics
VII} ed. D.~P. Landau, K.~K. Mon and H.~B. Sch\"uttler (Springer, 1994) 44.

\end{thebibliography}
\end{document}